\documentclass[sigconf]{acmart}
\AtBeginDocument{%
  }

\copyrightyear{2026}
\acmYear{2026}
\setcopyright{cc}
\setcctype{by}
\acmConference[CF Companion '26 ]{Proceedings of the 23rd ACM International Conference on Computing Frontiers}{May 19--21, 2026}{Catania, Italy}
\acmBooktitle{Proceedings of the 23rd ACM International Conference on Computing Frontiers (CF Companion '26 ), May 19--21, 2026, Catania, Italy}
\acmDOI{10.1145/3801488.3807497}
\acmISBN{979-8-4007-2569-2/2026/05}

\acmISBN{979-8-4007-2569-2/2026/05}

\usepackage{cancel}

\usepackage{acronym}
\usepackage{enumitem}

\usepackage{tikz}
\usetikzlibrary{positioning,arrows.meta,fit,calc,backgrounds}
\newacro{eol}[EOL]{End-of-Life}
\newacro{ict}[ICT]{Information and Communication Technology}
\newacroplural{ict}[ICT]{Information and Communication Technologies}
\newacro{lca}[LCA]{Life Cycle Assessment}
\newacro{lci}[LCI]{Life Cycle Inventory}
\newacro{pcr}[PCR]{Product Category Rules}
\newacro{epd}[EPD]{Environmental Product Declaration}

\usepackage{xcolor}
\colorlet{red}{black}
\begin{document}

\title{All LCA models are wrong. Are some of them useful?\\ Towards open computational LCA in ICT}

\author{Vincent Corlay}
\affiliation{%
  \institution{Mitsubishi Electric R\&D Centre Europe}
  \city{Rennes}
  \country{France}
}
\email{v.corlay@fr.merce.mee.com}

\author{David Bekri}
\affiliation{%
  \institution{iCoSys institute, HES-SO}
  \city{Fribourg}
  \country{Switzerland}
}
\email{david.bekri@hefr.ch}

\author{Marie-Anne Lacroix}
\affiliation{%
  \institution{Univ Rennes, CNRS, IRISA}
  \city{Rennes}
  \country{France}
}
\email{marie-anne.lacroix@irisa.fr}

\author{Maxime Pelcat}
\affiliation{%
  \institution{INSA Rennes}
  \city{Rennes}
  \country{France}
}
\email{maxime.pelcat@insa-rennes.fr}

\author{Maxime Péralta}
\affiliation{%
  \institution{Univ. Grenoble Alpes, CEA, List}
  \city{Grenoble}
  \country{France}
}
\email{maxime.peralta@cea.fr}

\author{Pierre-Yves Pichon}
\affiliation{%
  \institution{Mitsubishi Electric R\&D Centre Europe}
  \city{Rennes}
  \country{France}
}
\email{p.pichon@fr.merce.mee.com}

\author{Leo Saillenfest}
\affiliation{%
  \institution{Fraunhofer Institute for Reliability and Microintegration IZM}
  \city{Berlin}
  \country{Germany}
}
\email{leo.saillenfest@izm.fraunhofer.de}

\author{Olivier Weppe}
\affiliation{%
  \institution{INSA Rennes}
  \city{Rennes}
  \country{France}
}
\email{olivier.weppe@insa-rennes.fr}

\author{Sébastien Rumley}
\affiliation{%
  \institution{iCoSys institute, HES-SO}
  \city{Fribourg}
  \country{Switzerland}
}
\email{sebastien.rumley@hefr.ch}


\renewcommand{\shortauthors}{Corlay et al.}

\begin{abstract}
Life Cycle Assessment (LCA) is increasingly used to quantify and regulate the environmental impacts of Information and Communication Technology (ICT) systems.
Since direct biosphere measurements are complicated to perform, we claim that the environmental impact assessment of ICT relies heavily on models.
In this paper, we first revisit the fundamentals of LCA: we emphasize that ICT LCAs effectively form systems of models, and we argue that such systems require an “extra‑high” level of carefulness in construction, calibration, integration, and interpretation. We then document how this level of rigor is challenging to achieve with current practices. This is illustrated with emblematic examples of model misuse and an analysis of structural challenges related to database choice, scope mismatches, opaque aggregation, and model integration. From this analysis, we derive four key requirements for credible ICT LCA: explicit model lineage, clearly defined model scope, end‑to‑end traceability, and managed non‑obsolescence. Finally, we propose a framework that operationalizes these requirements using explicit dependency graphs, an open and versioned LCA-oriented model repository, automatic enforcement of integrity constraints, and a well‑defined model taxonomy. 
%We describe how these principles are applied as design rules in the new model‑oriented electronics LCA database ElecImpact.
\end{abstract}





\maketitle

%\tableofcontents

\section{Introduction}

%\subsection{Environmental costs of \acs{ict} and their evaluation with LCA}

 \begin{figure}[t]
\begin{tikzpicture}[
  font=\sffamily\small,
  node distance=6mm and 16mm,
  >=Latex,
  rounded corners=2pt,
  chall/.style = {draw, thick, fill=blue!6,    minimum width=30mm, align=left, inner sep=3pt},
  req/.style   = {draw, thick, fill=orange!18, minimum width=30mm, align=left, inner sep=3pt},
  sol/.style   = {draw, thick, fill=orange!2, minimum width=30mm, align=left, inner sep=3pt},
  head/.style  = {font=\bfseries\footnotesize, anchor=west},
  edge/.style  = {->, very thick, draw=black}, % all edges black
]

%--- column anchors (adjust 5.3 to bring columns closer if needed) ---
\coordinate (leftcol)  at (0,0);
\coordinate (rightcol) at ($(leftcol)+(5.3,0)$);

%--- headers ---
\node[head, font=\bfseries\large] (hL) at ($(leftcol)+(-1.1,2)$)  {LCA curses};
\node[head, font=\bfseries\large] (hR) at ($(rightcol)+(-1.6,2)$)   {LCA requirements};

%-----------------------
% Left column (challenges)
%-----------------------
\node[chall]                 (c1) at ($(leftcol)+(0,0.8)$) {Difficulty to list all\\ biosphere flows\\Section~\ref{sec_biosphere_flow}};
\node[chall, below=of c1]    (c2)                         {Curse 1: Model validation\\is difficult. Section~\ref{sec_valid_mod}};
\node[chall, below=of c2]    (c3)                         {Curse 2: Have to compose\\ models. Section~\ref{sec_curse_compose_model}};

%-----------------------
% Right column (requirements + solutions)
%-----------------------
\node[sol]                 (s1)  at ($(rightcol)+(0,1)$)  {Modeling approach\\Section~\ref{sec_modeling_approach}};

\node[req, below=4mm of s1] (rq1)                       {Model lineage \emph{(R1)}};
\node[req, below=2.5mm of rq1] (rq3)                    {Scope of models  \emph{(R2)}};
\node[req, below=2.5mm of rq3] (rq2)                    {Traceability  \emph{(R3)}};
\node[req, below=2.5mm of rq2] (rq4)                    {Non-obsolescence  \emph{(R4)}};

%-----------------------
% Edges (all black)
%-----------------------
\draw[edge] (c1.east) -- (s1.west);

\draw[edge, draw=orange!40] (s1.west) -- (c2.east);
\draw[edge, draw=orange!40] (s1.west) -- (c3.east);

\draw[edge] (c2.east) -- (rq1.west);
\draw[edge] (c2.east) -- (rq2.west);
\draw[edge] (c2.east) -- (rq3.west);
\draw[edge] (c2.east) -- (rq4.west);

\draw[edge] (c3.east) -- (rq1.west);
\draw[edge] (c3.east) -- (rq2.west);
\draw[edge] (c3.east) -- (rq3.west);
\draw[edge] (c3.east) -- (rq4.west);

% \draw[edge] (c7.east) -- (s1.west);
\end{tikzpicture}
%\vspace{-6mm}
\caption{LCA curses and corresponding requirements (Section~2).}
  \label{fig:lca-bipartite_requirements}
 % \vspace{-4mm}
\end{figure}
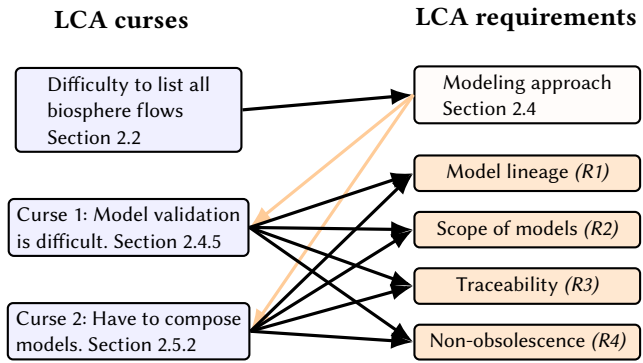

%%%FIG SOLUTIONS%%

\acf{ict} is clearly an industry with non-negligible environmental impacts.
%\py{if considered from the point of view of its infrastructure construction and maintenance.}.
%\mal{\acf{ict} is an environmentally harmful industry: ICT sector is estimated to represent between x\% and x\% of global carbon footprint, Distributed between user terminal (x\%), datacenters (x\%) and networks (x\%) [ref et chiffres à vérifier]. To meet the Paris Agreement, ICT sector should reduce its emissions by 45\% by 2030 \cite{SBTI2020}. But this is not the only environmental issue associated with ICT.}
%[\textcolor{red}{Comment: Le secteur des TIC représente environ 2 à 4 \% des émissions mondiales, dont les réseaux télécoms comptent pour un tiers. Les opérateurs doivent réduire de moitié leur empreinte carbone d’ici 2030 (SBTI [ref]).} \textcolor{blue}{David : à voir si redondant avec ce que j'ai rajouté ci-après]}. 
On one hand, ICT is  a voracious energy consumer: the International Energy Agency (IEA) stated in 2025 that the ICT sector consumed 1000 TWh of electricity in 2023, about 4\% of global electricity use \cite[Sec. 2.4]{IEA2025}, and the
%including around 25\% to 30\% attributable to networks \cite{IEA2025}, which aligns with previous predictions \cite{jones2018stop, Freitag2021}. 
 equivalent to Japan’s yearly electricity demand \cite{IEA2021JapanDOI}. 
 On the other hand, fabricating ICT equipment undeniably pollutes the environment, for instance by emitting CO2 (hundreds of millions of tons of CO2, expected to be more than 1\% of the total emissions by 2030 \cite[p. 17]{gesi2024digital}) and other greenhouse gases, or by depleting fresh water reserves (see \cite{wang2023environmental} for wafer production). 

 %Does that mean we must ditch \ac{ict} altogether? No, because it would create many other problems given the extent of the dependency of our modern societies on information technologies.
 However, abandoning ICT altogether is neither realistic nor desirable, given the deep reliance of modern societies on information technologies.
 %\py{proposition: Therefore it is becomes necessary for 'the society' (individuals, governments, organizations..) that the services provided by ICT are not outweighted by the impacts these services have on the natural environment. While such assessemnt requires a value judgement to compare the 'needs' and 'wants' of 'the society' with natural ecosystem services, thus poses fundamental ethical questions (which we do not aim at discussing within this paper), it also requires practical ways to evaluate how environmentally (un)sustainable these services and their underlying infrastructure are.}
 %\py{Or: However, it can offer services that may environmentally outperform more traditional ways the ‘needs’ and ‘wants’ of individuals and groups of people.}
 But individuals, organizations and societies have to begin to ask themselves \emph{``how much \ac{ict} is too much?"}.  Or, expressed more technically, \emph{when do the environmental costs associated with ICT outweigh its net contribution to social welfare?} 
This question lies at the heart of the notion of \emph{digital sufficiency} \cite{Santarius2022DigitalSufficienc}, which frames the need to keep ICT uses within levels compatible with environmental constraints and societal goals (as illustrated by Kate Raworth’s ``doughnut'' framework\cite{Raworth2017} and quantified by initiatives such as the SBTi targets \cite{SBTiInitiative}). %, as illustarted by “doughnut” of Kate Raworth frames the safe and just space where ICT delivers social value while respecting ecological ceilings and meeting social foundations.
To address this fundamental matter, individuals, organizations and societies \emph{need to be able to estimate the environmental cost of ICT} (and also the social cost, but that is out of scope of this contribution).
 
 Because calculating the ``net contribution to social welfare" is also challenging, and because not all entities are equally motivated to contribute to global social welfare, governments and regulators introduce limits, quotas and taxes to curb environmentally harmful industries \cite{UNEPreport2019, EUreview2025}. ICT will be no exception.
 %\seb{ici je me demande si c'est bien clair le raisonnement que les limites, quotas et taxes sont arbitraires parce qu'on ne sait pas trop comment calculer actuellement... mais ce n'est pas vital pour la structure du papier} 
 However, in order to tell where a company should stop producing or using ICT equipment, to ensure quotas are respected, and to calculate taxes, governments and regulators also \emph{need to be able to estimate the environmental cost of ICT}.
 
Finally, companies are accountable to shareholders and customers, who increasingly demand measurable environmental progress. Organizations are also encouraged by insurers to reduce their environmental impact. Thus, an increasing number of companies publish environmental reports detailing their cumulative negative impacts and outlining the steps they are taking to mitigate them. In addition, they actively seek environmental labels and certifications to demonstrate their commitment and strengthen their credibility. But to award labels, to calculate relative environmental improvements, again, they \emph{need to be able to estimate the environmental cost of ICT}.

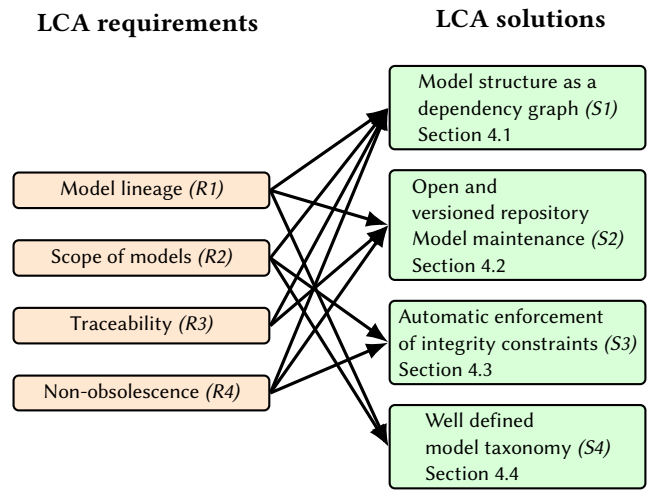
\begin{figure}[t]
\begin{tikzpicture}[
  font=\sffamily\small,
  node distance=6mm and 16mm,
  >=Latex,
  rounded corners=2pt,
  req/.style   = {draw, thick, fill=orange!18, minimum width=34mm, align=left, inner sep=3pt},
  sol/.style   = {draw, thick, fill=green!15,  minimum width=34mm, align=left, inner sep=3pt},
  head/.style  = {font=\bfseries\footnotesize, anchor=west},
  edge/.style  = {->, very thick, draw=black}, % all edges black
]

%--- column anchors (reduce 5.3 to bring columns closer if needed) ---
\coordinate (leftcol)  at (0,0);
\coordinate (rightcol) at ($(leftcol)+(5.0,0.0)$);

%--- headers ---
\node[head, font=\bfseries\large] (hL) at ($(leftcol)+(-1.5,2.5)$) {LCA requirements};
\node[head, font=\bfseries\large] (hR) at ($(rightcol)+(-1.2,2.6)$) {LCA solutions};

%-----------------------
% Left column (requirements)
%-----------------------
\node[req]                 (rq1) at ($(leftcol)+(0,0.3)$) {Model lineage \emph{(R1)}};
\node[req, below=4mm of rq1]   (rq3)                         {Scope of models \emph{(R2)}};
\node[req, below=4mm of rq3]   (rq2)                         {Traceability \emph{(R3)}};
\node[req, below=4mm of rq2]   (rq4)                         {Non-obsolescence \emph{(R4)}};

%-----------------------
% Right column (solutions – GREEN)
%-----------------------
\node[sol]                 (s1)  at ($(rightcol)+(0,1.4)$) {Model structure as a\\
 dependency graph \textit{(S1)}\\Section~\ref{sec_model_structure}};
\node[sol, below=2.5mm of s1] (s2)                          {Open and \\
versioned repository\\Model maintenance  \textit{(S2)} \\Section~\ref{sec_open_and_versioned_repo}};
\node[sol, below=2.5mm of s2] (s3)                          {Automatic enforcement\\ of integrity constraints \textit{(S3)}\\ Section~\ref{auto_enforc_inte_constraints}};
\node[sol, below=2.5mm of s3] (s4)                          {Well defined \\ model taxonomy \textit{(S4)}\\Section~\ref{sec_model_cate}};

%-----------------------
% Edges (requirements -> solutions; all black)
%-----------------------
\draw[edge] (rq1.east) -- (s1.west);
\draw[edge] (rq1.east) -- (s2.west);
\draw[edge] (rq1.east) -- (s4.west);

\draw[edge] (rq2.east) -- (s1.west);
\draw[edge] (rq2.east) -- (s2.west);
%\draw[edge] (rq2.east) -- (s3.west);

\draw[edge] (rq3.east) -- (s1.west);
\draw[edge] (rq3.east) -- (s3.west);
\draw[edge] (rq3.east) -- (s4.west);

\draw[edge] (rq4.east) -- (s1.west);
\draw[edge] (rq4.east) -- (s2.west);
\draw[edge] (rq4.east) -- (s3.west);

\end{tikzpicture}
\vspace{-3mm}
\caption{LCA requirements and proposed solutions (Section~4).}
  \label{fig:lca-bipartite_solutions}
    \vspace{-4mm}
\end{figure}

Environmental science experts have converged on standardized \ac{lca} methodologies to assess \emph{direct}\footnote{Indirect and systemic effects (for example rebound effects, induced demand, or long-term socio-technical transformations) remain difficult to quantify in a robust and comparable way. Nevertheless, recent works, including a report by ADEME \cite{ADEME2025_IT4Green_UseCases}, try to assess both direct and indirect effects for concrete digital use cases, using consequence trees and design-oriented approaches.} environmental impacts. %These methodologies have become the dominant approach for estimating the direct environmental cost of ICT \seb{petit sentiment de redondance entre les deux phrases précédentes, mais rien de très grave}. 
They \emph{aim to} quantify environmental impacts across the entire life cycle of an ICT system, following a so-called cradle-to-grave approach, from raw material extraction and manufacturing, through operation, to end-of-life treatment.

LCA methodologies \emph{aim to} estimate environmental costs and impacts. But \emph{aiming} is not \emph{succeeding}; being the dominant approach does not always imply it is a valid approach. How certain are we, as a community, that LCA methodologies truly capture the impact they intend to measure? That the numbers they provide can be trusted to \emph{estimate the environmental costs of ICT}, and this at the very moment we need them to calculate taxes or insurance fees, to introduce quotas and fix limits?

In this paper, we aim to reassure the reader: we believe that \ac{lca} methods are fundamentally sound; they have been successfully applied and can produce reliable results (\textcolor{red}{under strict conditions \cite{LUBECKI2025LCAReliability,Nordelof2014EnvironmentalImpactsElectrifiedVehicles}}). However, they must be used appropriately and interpreted with full awareness of their limitations. One of the most significant limitations is that, when applied to complex systems, LCA theoretically requires a divide-and-conquer approach extending to the most elementary components of the system. In practice, this quickly becomes intractable (e.g., how many elementary components are there in a laptop, and what exactly qualifies as an elementary component?). Consequently, one must rely on models to make the analysis feasible. For instance, rather than explicitly modeling every via, trace, and solder joint on a printed circuit board (PCB) and associated production processes, one may rely on a parametric PCB model that estimates environmental impacts from a limited set of design parameters, such as board area, number of layers, substrate type, and surface‑mount component density.

However, the introduction of models inevitably brings additional uncertainty and potential error. By definition, models simplify reality and therefore cannot perfectly represent it. They rely on hypotheses and assumptions that shape, and sometimes constrain, the results. The difficulty is amplified in the case of ICT equipment, which constitutes highly complex systems in their own right. Conducting an LCA of a device like a smartphone therefore requires not a single model but an interconnected set of models: a genuine \emph{system of models}. For such a \emph{system} to yield meaningful and robust results, its components must be tightly coupled and its overall complexity carefully controlled. By resorting to modeling, we avoid tedious successive decompositions down to biosphere flows.
%\mprl{la notion de flux élémentaires n'a pas été introduite à ce stage. Est-ce gênant ?}. 
Yet this simplification is not without trade-offs. Complexity does not disappear; it is displaced. The burden shifts from exhaustive structural decomposition to the \emph{careful construction, calibration, integration, and interpretation} of models.

We co-authors have doubts that we have these systems of models under control. We doubt that we \emph{construct, calibrate, integrate and interpret} models with the required care. If we do not, then our confidence in the correct application of LCA methods may be largely unwarranted. 
We are not alone in raising this concern. A recent review  \cite{KamiyaCoroama2025DataCentreEnergy} by the IEA’s 4E Technology Collaboration Programme (4E TCP), which analyzes over 100 publications estimating data‑center energy use, highlights two key observations: First, these estimates rely almost exclusively on models\footnote{They identify the following main categories of models used in the literature for energy use estimation:
\emph{(i)} Bottom-up models: based on estimates of the installed server and IT equipment base, combined with equipment specifications (e.g., average server power consumption), equipment lifespans, and other energy-influencing attributes such as power-usage effectiveness.
\emph{(ii)} Aggregated totals: often described in the literature as \emph{top-down} approaches, relying on national, regional, or organizational energy consumption data that are measured or estimated at an aggregate level.
\emph{(iii)} Temporal proxy extrapolation: starting from an initial base estimate obtained using one of the methods above, this approach combines high-level proxies and indicators (e.g., data traffic or energy-intensity assumptions) to extrapolate data-centre energy use under varying activity and efficiency-improvement scenarios.}. Second, the resulting estimates exhibit very high variability across studies.

George Box famously wrote in 1976 that \emph{all models are wrong, some of them are useful} \cite{Box01121976}. In doing so, he acknowledged not only the unavoidable inaccuracy of models, but also a more uncomfortable implication: usefulness is not automatic. Some models are simply not useful \emph{and shall not be used}. 
This leads us to a deliberately provocative question: could it be that a substantial portion, perhaps even the majority, of our efforts to model the environmental impacts of ICT systems is \emph{not useful}? And \emph{shall not be used?}
 
This paper is a call to reconsider how we are using and applying LCA methodologies and models in the ICT sector. If we, the sustainable-ICT research community, want to estimate correctly the environmental figures society awaits us to deliver, we must be more careful about how we use them \textcolor{red}{(see Section~3 for examples)}.

To the extent of our knowledge, this paper is the first to pinpoint and classify the requirements and computational limitations of ICT \ac{lca} as practiced today. We also propose solutions to overcome these limitations \textcolor{red}{with a more structured \textit{computational-oriented} approach}.
We believe such solutions are crucial to build a bridge between environmental sciences and ICT, and to lay solid foundations for digital systems ecodesign.

This paper is organized in three parts. 
In Section~\ref{sec_challenges}, we go back to the fundamentals of  LCA \textcolor{red}{and good scientific practice}. We begin by documenting how \ac{lca} has progressively become embedded in regulations and legislations (Subsection~\ref{sec_LCA_redul}). In Subsection~\ref{sec_biosphere_flow}, we remind that LCA fundamentally consists of listing all biosphere flows, i.e., counting what each process takes from and returns to nature. We then discuss the difficulty of establishing this listing through direct measurements, notably because it means instrumenting factories, which is very intrusive (Subsection~\ref{sec_list_bio}). Because of this difficulty, we posit that LCA in ICT has essentially become a modeling activity, and so we review the principles of modeling in Subsection~\ref{sec_modeling_approach}. We exhibit in Subsection~\ref{sec_curses} that LCA in ICT suffers two curses that make it difficult to follow the canons of modeling: 1) models are intrinsically hard to validate, and 2) we have to compose models.
%\mprl{Sur la figure 1 il y a trois curse, et la troisième n'est pas décrite de la même manière que dans cette phrase}. 
\textcolor{red}{In Section~\ref{sec_uncertainty_practice}, we review methods that can \emph{quantify} some categories of uncertainty in LCA, while noting that such methods do not necessarily \emph{reduce} uncertainty itself.}
Finally, we deduce from all previous subsections that LCA in ICT requires an extra-high level of carefulness in the way models are addressed. We propose a set of requirements \emph{(R1--R4)} and list their benefits \emph{(B1--B8)} in Subsection~\ref{sec_requirements}.
%that LCA in ICT requires an extra-high level of carefulness in the way models and their lineage are documented.

In Section~3, we begin by listing (Subsection~\ref{sec_pb_model}) several instances where this “extra‑high level of carefulness” is absent, and we enumerate classes of challenges in current LCA practice in Subsection~\ref{sec_ex_LCA}. We map each example and each structural challenge to the relevant requirements \emph{(R1--R4)} and benefits \emph{(B1--B8)}.
We then discuss the underlying causes, “the why” \textcolor{red}{of natural model misuse} (Subsection~\ref{sec_natural_model_misuse}).

Section~4 finally introduces a framework that operationalizes the requirements identified earlier by \emph{(i)} organizing models as an explicit dependency graph (Section~\ref{sec_model_structure}), \emph{(ii)} embedding them in an open and versioned repository (Section~\ref{sec_open_and_versioned_repo}), \emph{(iii)} automatically enforcing integrity constraints (Section~\ref{auto_enforc_inte_constraints}), and \emph{(iv)} establishing a well-defined model taxonomy (Section~\ref{sec_model_cate}).

Figures~\ref{fig:lca-bipartite_requirements} and~\ref{fig:lca-bipartite_solutions} provide a structured overview of how current problems relate to the requirements and to the proposed solutions.

\section{Model-based LCA: needs,\\ definitions, design, curses, and requirements}
\label{sec_challenges}
%why we need transparency; LCA requirements
%LCA challenges

% \textcolor{red}{MPEL: Il n'est pas très clair pour moi ce qui relève des challenges (ce qui est difficile) et des requirements (quelles sont les conditions de réussite)}
\ac{lca} is the standard methodology used to assess the environmental impacts of products and services, as standardized by ISO $14040$–$44$ \cite{iso_14040, iso_14044}. It can either follow a cradle‑to‑grave approach in order to cover the entire life cycle of a product from raw material extraction to end‑of‑life; or can focus on a part of the life cycle of a product such as the production phase, following for example a cradle-to-gate approach.

\ac{lca} provides a comprehensive evaluation across multiple environmental impact categories, such as the 16 midpoint impact categories defined in the Environmental Footprint (EF) 3.1 method \cite{andreasi2023updated}, such as climate change (GWP), water use, land use, and minerals and metals use (ADP).
%Product Category Rules (PCR) and Product‑Specific Rules (PSR) \cite{iso_14025, WTO_rules} are developed to provide guidelines on how to conduct an \ac{lca} for specific categories of products, with the objective of improving consistency and comparability across studies.
%% Paragraphe plutôt pour partie régulation

\subsection{LCA as a regulatory support tool}
\label{sec_LCA_redul}

%\seb{Rappeler que LCA commence a avoir "force légale", reprendre les points de l'introduction.}

LCA is becoming increasingly critical because its use now extends beyond eco‑design and internal decision‑making to the regulatory domain. In addition to guiding designers toward more environmentally efficient solutions, LCA results are expected to support both reporting at the product leveland at the corporate-level - GHG Protocol standards (scopes 1/2/3) or the Corporate Sustainability Reporting Directive (CSRD, Directive (EU) 2022/2464). As a first step, Product Carbon Footprints (PCFs), built on top on ISO 14040/14044 LCA methodologies and focusing on Carbon Emissions only \cite{iso14067-2018}, are starting to be adopted by the industry, and can be reported in an \ac{epd}. Product Environmental Footprints (PEFs), for their part, are similar to PCFs but focus on a broader set of environmental criteria beyond carbon emissions (typically EF 3.1 categories). To facilitate and harmonize reporting, technical committees are currently drafting several \ac{pcr} for PCFs (PCR for PCFs) and Product Environmental Footprint Category Rules (PEFCRs), i.e., LCA methodologies specifying the scope, boundary conditions, functional units and other requirements for certain product categories. 

While PCFs are a first step for using LCA methodologies for reporting, the EU Commission is favouring the multi‑environmental ‑criteria solution (PEF), with the objective of supporting its regulatory framework, the Ecodesign for Sustainable Products Regulation (ESPR, Regulation (EU) 2024/1781). This framework will include the Digital Product Passport (DPP), an information system to disclose and manage product data over time. The DPP may for instance reference EPDs. 
Its technical implementation (data storage, archiving, persistence, reliability, integrity, interoperability, access rights management, APIs, etc.) is being supported by European standards developed by the CEN‑CENELEC JTC24 in response to a standardisation request from the European Commission.

As a result, the European Commission is expected to issue legislative acts requiring the publication of Product Environmental Footprints (PEF) based on LCA and setting minimum environmental criteria for ICT products placed on the EU market.

\subsection{Theoretical LCA: listing all biosphere flows}
\label{sec_biosphere_flow}

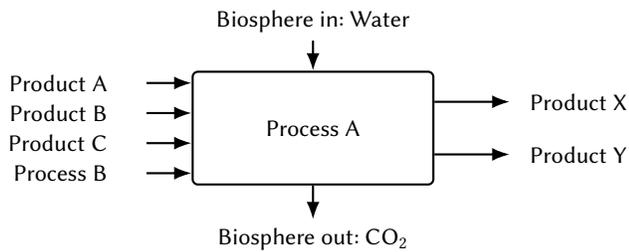
\begin{figure}
  \centering
\begin{tikzpicture}[
  >=Latex,
  thick,
  font=\sffamily,
  % ↓ smaller minimum size and padding
  proc/.style = {draw, fill=white, rounded corners=2pt,
                 minimum width=3.2cm, minimum height=1.5cm,
                 inner sep=2pt, align=center},
  inlab/.style = {anchor=east},
  outlab/.style = {anchor=west}
]
% Process box
\node[proc] (P) {Process A};

% Left inputs
\foreach \y/\txt in {0.6/{Product A}, 0.2/{Product B}, -0.2/{Product C}, -0.6/{Process B}}{
  \node[inlab] at ($(P.west)+(-1.0,\y)$) {\txt};
  \draw[->]     ($(P.west)+(-0.6,\y)$) -- ($(P.west)+(0,\y)$);
}

% Right outputs
\foreach \y/\txt in {0.35/{Product X}, -0.35/{Product Y}}{
  \draw[->] ($(P.east)+(0,\y)$) -- ($(P.east)+(1.0,\y)$);
  \node[outlab] at ($(P.east)+(1.15,\y)$) {\txt};
}

% Biosphere in/out
\draw[->] ($(P.north)+(0,0.4)$) -- (P.north);
\node[anchor=south] at ($(P.north)+(0,0.4)$) {Biosphere in: Water};

\draw[->] (P.south) -- ($(P.south)+(0,-0.45)$);
\node[anchor=north] at ($(P.south)+(0,-0.45)$) {Biosphere out: CO$_2$};
\end{tikzpicture}
\vspace{-5mm}
  \caption{Illustration of an LCA process with input/output technosphere (\textcolor{red}{products in the figure}) and biosphere flows.}
  \label{fig:process}
    \vspace{-3mm}
\end{figure}

A fundamental component of \ac{lca} is the \ac{lci}. It consists of listing all processes involved in the system, each process taking as input technosphere and biosphere flows and producing technosphere and biosphere flows, as illustrated in Figure~\ref{fig:process}. Technosphere flows primarily act as proxies to connect processes within the product system, while the ultimate objective of the inventory is to account for all relevant biosphere flows. In the theoretical formulation of \ac{lca}, this would require an exhaustive and explicit representation of all such flows across the entire life cycle.

\subsection{The difficulty of listing biosphere flows}
\label{sec_list_bio}

Measurement issues in \ac{lca} differ fundamentally from those encountered in experimental sciences such as physics. 
\begin{enumerate}[leftmargin=*, label=\textbf{\arabic*.}, labelsep=.5em, nosep]
\item Environmental burdens cannot be inferred simply from inspection of the final product, as many impacts occur upstream throughout the production chain during manufacturing processes, which themselves often cause material losses. 
% NOTE POUR LE JOURNAL: FAIRE LE LIEN AVEC LE NUTRISCORE, PARLER DE MARQUEURS
%are generated upstream and dissipated along complex supply and construction chains, 
This makes impacts often far remote in space and time from the finished product. 
\item Even when measurements are attempted during production, many relevant flows are intrinsically difficult to quantify, such as diffuse gaseous emissions or small material losses that are hard to capture with sufficient accuracy. 
\item Meaningful measurements have to be performed directly within industrial facilities, which is operationally complex, costly, and rarely compatible with routine manufacturing constraints. Moreover, this raises the issue of allocating biosphere flows to the specific product under study when production chains manufacture multiple devices. This challenge naturally leads to the introduction of \emph{models}, which are discussed in the following subsection.
\item Access to such measurements is further limited by confidentiality and trade‑secret concerns, which restrict the disclosure of detailed process data. 
\item Even if every flow were perfectly measurable, the theoretical number of flows involved in an exhaustive life‑cycle inventory would be intractable. For example to produce an electronic device, we should take into account the use of machines that enabled its manufacture, but also a certain proportion (again raising the allocation problem mentioned in \emph{(3)}) of the process that enabled this machine manufacture, and so on. Therefore, an \ac{lca} always begins with the definition of system boundaries (which processes are included in the assessment or not). In addition, \emph{cut-off criteria} are specified to justify the exclusion of processes or flows expected to be negligible, typically using thresholds defined by a norm or based on assessments. 
\item Finally, in contrast to physics, where measurements are typically performed on closed or controlled systems and model validation can rely on repeatable experiments with clearly defined hypotheses, LCA operates on open, evolving systems whose conditions are hard to recreate ex post; hence, re‑measurement and experimental replication are challenging, which further constrains model validation.
\end{enumerate}

As a consequence, direct measurement of biosphere flows is rarely feasible in practice, and environmental impacts must instead be inferred through models that combine heterogeneous data, assumptions, and proxies. \textcolor{red}{As an illustration, the notion of \textit{simplified} LCA is frequently adopted}. 

\subsection{LCA as a modeling activity}
\label{sec_modeling_approach}
%In electronics, for instance, the carbon footprint of a printed circuit board (PCB) cannot be read off the board itself, but must \mprl{can?} be inferred from design and manufacturing parameters such as the number of layers \mprl{area and manufaturing location?} \cite{LeGargasson2025PCBnCO}.
\subsubsection{Principle}
When it is not possible to explicitly list all biosphere flows, practitioners instead measure observable quantities, such as the weight of a smartphone, its price, its RAM, the number of layers of a PCB \cite{LeGargasson2025PCBnCO}, and use these measurements to \emph{infer}, rather than directly measure, environmental impacts. This inference relies on formalized procedures that relate observable product characteristics to unobservable environmental metrics. These procedures are what we refer to as models. Technical characteristics are thus often used as proxies for environmental exchanges, either by being converted into quantities of biosphere flows or by serving as inputs to background impact models\footnote{
%\seb{le footnote est trop compliqué pour moi... est-il vraiment nécessaire?}
We note that several notions commonly used in LCA, such as PCRs, foreground and background systems, primary and secondary data, and models, refer to distinct but interrelated concepts. 
%Confusing these notions can lead to methodological ambiguities, for instance by treating methodological rules as models.
%, or by conflating data sources with the models that process them. 
PCRs
%and PEF Category Rules (PEFCRs) 
define how an assessment should be conducted for a given product category, but they neither provide data nor constitute models themselves. 
An LCA is implemented through models related to the foreground system (which are specific to and often controlled by the studied system) with models related to background systems (outside the direct control of the studies). Models in both foreground and background systems may be calibrated based on primary data (direct measurement for the system under study) or secondary data (external sources). %, which rely on generic representations of upstream and downstream activities. 
%These models are parameterized using primary data, collected directly for the system under study, and secondary data, reused from databases or literature. 
%Understanding the distinction between these notions is essential to correctly interpret LCA results, assess their validity, and identify where assumptions, uncertainties, and limitations originate.
} 
(see Section~\ref{sec_model_structure} for examples of model categories). 

This turns effectively LCA into a \emph{modeling} activity.

\subsubsection{From ``wrong" to useful}

As mentioned in the introduction, all models are wrong. 
Nevertheless, being “wrong” does not always imply being useless. Newtonian mechanics is formally incorrect when compared to relativity or quantum mechanics, yet it remains sufficiently accurate within a well-defined domain of validity to support most engineering applications. Engineers continue to rely on such models because they are useful answers to specific questions, given acceptable error margins. %This consideration is tightly linked to the notion the scope definition and model validation. 

What we need to know is ``how wrong?", whether they are useful for the specific questions we seek to answer, and whether we can make them more useful through validation and transparency.

\subsubsection{Model design}
\label{sec_model_design}

Two main approaches can be distinguished when designing \ac{lca} models. 

\emph{(i) Data-driven or empirical modeling} relies on direct measurements of biosphere flows (primary data) obtained under specific experimental or industrial conditions, often costly, difficult to perform, and limited in scope as previously mentioned in Subsection~\ref{sec_list_bio}. These measurements are then used to construct empirical relationships between observable product or process characteristics and environmental exchanges, for instance through linear or polynomial regressions, or more complex techniques such as neural networks. 
% POUR LE JOURNAL: fournir référence
While such models can be effective within the range of observed data, they are exposed to well-known risks: they may fail when extrapolated beyond the calibration domain, they can overfit noisy or sparse data, and they often exhibit limited interpretability, making it difficult to assess their assumptions or domain of validity.

\emph{(ii) Physics-based or ``first-principles" modeling (axiomatic)} instead derives environmental exchanges from descriptions grounded in physical, chemical, or thermodynamic laws. These models are typically more interpretable and better suited for extrapolation. 
\textcolor{red}{The number of free parameters that must be specified or calibrated is usually reduced compared with purely empirical approaches. 
Nevertheless, this approach introduces the risk of \emph{model mismatch}: if the assumed model structure, namely the set of governing equations, boundary conditions, and simplifying assumptions used to represent the system, is incomplete or inaccurate \cite{ShlezingerEldar2023ModelBasedDL}, the constrained parameterization may bias inferred environmental exchanges and impacts (see also Section~\ref{sec_uncertainty_practice}).}

In practice, \ac{lca} studies often combine both approaches. %, \seb{la phrase suivante je comprends pas trop} embedding empirical sub-models within broader physics-based frameworks, or conversely constraining data-driven models using physical consistency requirements (e.g. by mass balance between inputs and outputs). 

\subsubsection{Model composition}
\label{sec_mod_combin}

Beyond empirical and physics‑based designs, practitioners can create new models by composing existing ones: an effective divide‑and‑conquer for high-complexity systems. However, composition comes with strict guardrails: \emph{(i)} constituent sub‑models must be validated (see the following Subsection~\ref{sec_valid_mod}) within their respective domains; \emph{(ii)} their assumptions must be mutually compatible (units, scopes, boundary conditions, allocation choices, operating regimes) and integration can induce second‑order interactions (feedbacks, cross‑terms, double counting) that were absent or negligible at the unit level. \emph{(iii)} Crucially, invalidation must propagate: if any sub‑model is later found erroneous (e.g., due to a defective probe or flawed calibration), the composite model inherits that invalidation, at least for the portions that depend on the faulty component. Ideally, composite models would themselves be validated as valid composites according to some rules. 
 
\subsubsection{Model validation is necessary}
\label{sec_valid_mod}

Regardless of the modeling approach adopted, validation is indispensable. 

\emph{(i)} In both empirical and physics-based approaches, a model must ultimately be confronted with observations (preferably not the ones used to train the models) to establish its credibility. \textcolor{red}{This is a key principle of the scientific method.} Even a theory as foundational as the one presented in \emph{Philosophiæ Naturalis Principia Mathematica} by Isaac Newton derived its scientific legitimacy from systematic comparison between predictions and measured phenomena; anchoring a model to reality therefore inevitably requires measurement. 

\emph{(ii)} In some cases, however, models are proposed before such validation can be fully achieved. These models must then be regarded as \emph{conjectural}: they may provide useful insights or support exploratory analyses, but they remain exposed to potential invalidation by future experimental evidence.

\subsection{The curses of model-based LCA}
\label{sec_curses}
%\mprl{un tout petit peut redondant avec les deux sections précedentes, non ? Ca vaudrait pas le coup de transférer les éléments nouveaux de cette section dans les sections d'avant pour la transformer en un genre de "wrap-up" ? Ou alors de tout fusionner ?}

Summarizing Subsections~\ref{sec_biosphere_flow}, \ref{sec_list_bio}, and \ref{sec_modeling_approach}, we posit that LCA in ICT is subject to the two following curses.
%The initial problem of listing all biosphere flows (Section~\ref{sec_list_bio}) is now shifted to the following curses in model-based LCA.

\subsubsection{Curse 1: Model validation is difficult}

%As detailed in Subsection~\ref{sec_valid_mod}, the validity of an \ac{lca} model can only be established either through direct confrontation with measurements, when they exist, or through explicit dependencies on other models that have themselves been validated. 

As discussed in Section~\ref{sec_list_bio}, measurement in LCA is highly challenging. As a result, assessing the validity of an \ac{lca} model is both intrinsically difficult and very costly. Even for seemingly simple objects, such as a single screw or an electronic component, directly validating predicted environmental impacts through measurements is often impractical and this difficulty increases rapidly with system complexity.%, for instance when considering enclosures, printed circuit boards, or complete devices.

Because comprehensive validation is so resource‑intensive, each direct measurement of biosphere flows (typically primary data) has an exceptionally high value and must therefore be produced, documented, and referenced with extreme care. 

\subsubsection{Curse 2: We have to compose models}
\label{sec_curse_compose_model}

Because of the aforementioned complexity of ICT systems, we cannot avoid composing models. Nevertheless, when composing models in LCA, it is even less likely to have access to measurements. %Therefore, to validate those composite models, all the hypotheses made have to be valid for all sub-models involved. This highlights the importance of being able to quantify the degree of uncertainty of estimations and carefully define the domain of validity. 
Moreover, not all models evolve at the same pace in LCA: some are regularly updated to reflect new knowledge or technologies, while others remain unchanged for long periods of time. In certain cases, several competing models coexist \textcolor{red}{(see examples in Section~\ref{sec_ex_LCA})}. As a result, the requirements of Section~\ref{sec_mod_combin} are even more relevant.

For instance, when performing an LCA with models based on secondary data, ensuring consistency with fundamental conservation principles, e.g., energy conservation (first law of thermodynamics) and mass balance, is challenging.
% when tracing material and energy flows.
%Other resources accounting schemes, such as MFA (Material Flow Analysis) may be more suited for this task and would benefit from a coupling with LCA models.

\subsection{\textcolor{red}{LCA uncertainty assessment}}
\label{sec_uncertainty_practice}
%Uncertainty is an inherent feature of \ac{lca} results and arises from imperfect knowledge of inventory data, methodological choices, and scenario assumptions.
\textcolor{red}{The observations outlined above make \textit{uncertainty} an inherent feature of \ac{lca} results. Different categories of uncertainty have long been identified in the LCA literature \cite{HeijungsHuijbregts2004LCAuncertainty}\footnote{The authors also highlighted the limited consideration of uncertainty in many LCA studies, noting that “it is amazing that this interest has not been natural since the development of LCA and the rise of its use”.}. For instance, one can distinguish \textit{parameter uncertainty}, which relates to uncertainty in numerical input values, from \textit{model uncertainty} (closely related to model mismatch as discussed in Section~\ref{sec_model_design}), which arises from methodological choices.  Contrary to parameter uncertainty, model uncertainty cannot always be addressed through probabilistic parameter variation alone.}

\textcolor{red}{The uncertainty assessment methods recalled below, also relying on explicit modeling assumptions, aim to \textit{quantify} uncertainty. The requirements proposed in the following Section~\ref{sec_requirements} go beyond quantification: they also aim to \textit{mitigate} uncertainty when possible and, more generally, to limit model misuse through the adoption of adapted practices.}

\textbf{\textcolor{red}{Uncertainty assessment methods}}
%\label{subsec_uncert_assess}
\textcolor{red}{Parameter uncertainty is usually modeled by assigning a probability distribution to a parameter. It may pertain either to the data itself or to the mapping of a secondary dataset (see ``scope mismatches" in Section~\ref{sec_ex_LCA}), itself having a potential uncertainty. The distributions are then combined to model the resulting uncertainty.}

\textcolor{red}{When measurement data or statistical evidence are unavailable, the probability distribution may be inferred through a \textit{model}.
A widely used model is the PEDIGREE matrix, where an expert choses scores instead of an explicit distribution. Initially proposed by Weidema and Wesn{\ae}s \cite{WeidemaWesnaes1996pedigree}, this method translates \textit{qualitative data quality indicators}\footnote{Inventory data are evaluated according to criteria such as reliability, completeness, temporal correlation, geographical correlation, and technological correlation. Each criterion is scored, and the scores are mapped to uncertainty factors, which are subsequently combined to derive a probability distribution for each parameter, most often assumed to be lognormal.} into \textit{quantitative uncertainty estimates}, i.e., a probability distribution.
The PEDIGREE approach is implemented in mainstream \ac{lca} databases and softwares \cite{Weidema2013ecoinventDQ,Ciroth2016EmpiricalPedigree}.}
\textcolor{red}{We emphasize that the PEDIGREE approach is itself a modeling layer and therefore entails assumptions, model error, and potential out‑of‑scope use. Ecoinvent acknowledges that this model requires continued improvement and validation, including comparisons with uncertainty estimated from empirical data \cite[Chap.~10]{Weidema2013ecoinventDQ,Ciroth2012Pedigree}.}  

\textcolor{red}{
Alternatively, Bayesian methods, unlike the PEDIGREE approach, allow uncertainty to be directly informed and updated using new measurement data or alternative data source \cite{Ascough2008BayesianLCA}.}

\subsection{Induced requirements and their benefits}
\label{sec_requirements}
We deduce that the curses of Section~\ref{sec_curses} induce the following requirements. \textcolor{red}{Figure~\ref{fig:lca-bipartite_requirements} illustrates the connections between the curses and the proposed requirements}. For each requirement, we discuss the benefits associated with its satisfaction.

\textbf{Model lineage \emph{(R1)}.}
\label{sec_req_model_lineage}
Each model should come with an explicit description of its ancestry (the sub-models it relies upon) and siblings (the models that use it) and dependencies.
This includes both its links to empirical measurements (when they exist) and its links to upstream models from which it inherits assumptions or structure.
In other words, we should be able to distinguish models that have “royal blood” (i.e. a clear chain back to observations) from models that rest on conjectural or weakly justified assumptions.\\
\underline{Benefit}: Making this lineage explicit allows practitioners to assess how much empirical support underpins a given result \emph{(B1)} and to identify where uncertainty or speculation enters the chain \emph{(B2)}.
It also helps prevent \emph{model washing} \emph{(B3)}, whereby a weakly supported or poorly validated model is embedded within, and thereby legitimized by a more reputable model.

\textbf{Scope of models \emph{(R2)}.}
\label{sec_scope_models}
A model is never valid “in general”; it is valid only within a specific domain defined by hypotheses on units, system boundaries, operating conditions, and technological context.
Composing models therefore requires checking that their scopes are compatible: a model that outputs a quantity in one unit cannot be used as if it directly provided another (e.g. treating $1$\,L of water as $1$\,kg at $1$\,bar and $4^\circ$C), and simplifying assumptions must not be silently violated in downstream uses.\\
\underline{Benefit}: By making scope definitions explicit, practitioners can identify incompatible assumptions early, preventing composite models from appearing coherent while resting on inconsistent premises \emph{(B4)}. It also prevents out-of-domain reuse \emph{(B5)}.
%Without explicit scope definitions, composite models may appear coherent while in fact resting on mutually incompatible assumptions.

\textbf{Traceability \emph{(R3)}.}
\label{sec_trac_scope}
When a model is used, it should be possible to explain what it does, how it was built, and how a given numerical result was obtained.
In practice, this means that calculations can be “unwound” into a dependency tree that closely resembles the idealized process graph introduced earlier. 
For example, when reporting an aggregate figure such as “data centres consume on the order of 400~TWh of electricity per year” \cite{IEA2025,IEA_DataCentres_2025}, it should in principle be possible to recompute this value end‑to‑end from documented data, code, and modeling choices, ideally made available in a public, versioned repository.
Note that several scientific papers do publish their inventories and models in an open‑source manner; see, for instance, \cite{Loubet2023ICT_HE, Zhang2022HV_AEC_LCA, Zhang2023AEC_ManufacturerPerspective, Krishnan2008HybridLCI_Semiconductors, Nordelof2018EV_Motor_LCI_PartI, Nordelof2019EV_Inverter_LCI_PartI, Nordelof2019EV_Inverter_LCI_PartII,Falk2025MoreThanCarbon}.\\
\underline{Benefit}: Traceability thus connects impact numbers back to models \emph{(B6)} and, ultimately, to observations (also through lineage) \emph{(B1)}. It supports explanation, replication, and auditing of results \emph{(B7)}.

\textbf{Non‑obsolescence \emph{(R4)}.}
\label{sec_model_maintenance}
Finally, models and databases evolve over time as new data, technologies, and methodological insights appear.
A credible framework must therefore organize this evolution rather than ignore it. \\
\underline{Benefit}: Non‑obsolescence preserves the interpretability of published results in light of current knowledge: past model versions stay archived and citable \emph{(B6)}, but it is also possible to identify when a result depends on outdated components and how it would change under updated models \emph{(B8, B4)}.

\textcolor{red}{In substance, requirements \emph{(R1--R4)} articulate what \textit{good scientific practice} demands.}
%of serious modeling. A well‑conducted \ac{lca} would ideally satisfy these requirements.} \seb{je suis pas sur que ce soit nécessaire d'ajouter cette phrase. On pourrait nous reprocher de définir nous ce qu'est un bon modèles.. on prend moins de risque à simplement énumérer les défauts...}

\section{Patterns and causes of reduced methodological rigor with respect to the requirements}
This section illustrates how methodological rigor is often relaxed in practice when using models and why this happens, even in good faith.
We first present a few emblematic examples of model misuse (both within and outside the field of LCA), then summarize recurring structural challenges in current LCA practice, and finally discuss why these misuses are “natural” in the absence of the aforementioned explicit safeguards: model lineage \emph{(R1)}, scope \emph{(R2)}, traceability \emph{(R3)}, and non‑obsolescence \emph{(R4)}.

\subsection{Illustrative example of model abuse}
\label{sec_pb_model}

\textbf{Reinhart and Rogoff.}
In \cite{ReinhartRogoff2010}, Reinhart and Rogoff reported that public debt levels above 90\% of GDP are associated with strongly reduced or negative growth.
This finding was quickly interpreted as a quasi‑rule justifying strict austerity policies.
Herndon et al.~\cite{HerndonAshPollin2014} later uncovered spreadsheet errors and questionable methodological choices that materially affected the results, but by then the original claim had already influenced policy.
The failure did not stem solely from a flawed calculation; it came from elevating a fragile empirical pattern to a law, ignoring uncertainty and context.\\
\underline{Possible improvement}: \emph{(B2)} via \emph{(R1)} – ``Where does uncertainty/ speculation enter?'', \emph{(B7)} via \emph{(R3)} – ``Supporting explanation, replication, and auditing''.

\textbf{Gaussian copulas.}
Gaussian copula models were widely used to price complex credit products such as CDOs.
While mathematically coherent, they relied on assumptions of relatively stable correlations and limited tail dependence.
In practice, they were applied beyond their validity domain, including for systemic risk assessment.
When correlations surged during the 2007–2008 crisis, risks were underestimated:
the issue was less the formula itself than the uncritical reuse of a convenient model outside the conditions under which it had been validated \cite{MacKenzieSpears2014}.\\
\underline{Possible improvement}: \emph{(B2)} via \emph{(R1)} – ``Where does uncertainty/ speculation enter?'', \emph{(B5)} via \emph{(R2)} – ``Preventing out‑of‑domain reuse''.

\textbf{Footprint of an email.}
Popular figures such as “4–50\,g\,CO\textsubscript{2} per email” conflated  marginal and average impacts by mixing the fixed energy consumption of devices, data centers, and networks with the incremental cost of sending a single message.
Later clarifications emphasized that, in most realistic contexts, the marginal footprint of an attachment‑free email is orders of magnitude smaller than these headline numbers \cite{BernersLeeHowBadBananas2010,BernersLeeHowBadBananas2020,CarbonLiteracyEmail,MobileSyrupEmailMyth}.
Mike Berners-Lee posted the following in 2020 on Twitter: ``To clarify, following FT and BBC pieces, the carbon footprint of sending an email is trivial. Looks like UK gov has misused  a press release from OVO that in turn used  estimates from the 2010 version of my book 'How Bad Are Bananas?' (now updated)."
Here, a rough illustrative model was reinterpreted as a precise and context‑independent fact, despite its correction.
\underline{Possible improvement}: \emph{(B2)} via \emph{(R1)} – ``Where does uncertainty/ speculation enter?'',  \emph{(B5)} via \emph{(R2)} – ``Preventing out‑of‑domain reuse''.

\textbf{Ecologits.}
Ecologits \cite{EcoLogits2024} is an environmental footprint calculator for AI models.
Even with a documented methodology, it is practically impossible to track all hardware actually used for inference. 
More critically, during the communication of results based on this tool by some authors of this paper \cite{aidays}, the calculator and its underlying database were updated, significantly changing the impacts. 
This situation raises several issues: \emph{(i)} a lack of transparency regarding updates to the database and the parameters used in the calculations; \emph{(ii)} the impossibility of accessing previous versions of the database in order to reproduce or compare past results; and \emph{(iii)} the absence of traceability of dependencies and of the downstream effects that such updates may have on existing results or on other models relying on this calculator.\\
\underline{Possible improvement}: \emph{(B3)} via \emph{(R1)} – ``Avoiding model washing'' (by preventing potentially weakly supported sub‑models used by those trusted databases from being legitimized simply because they are embedded),
\emph{(B6)} via \emph{(R3)} – ``Connecting numbers back to models and data'' \emph{(B7)} via \emph{(R3)} – ``Supporting explanation, replication, and auditing'', \emph{(B8)} via \emph{(R4)} – ``Handling outdated components and evolution over time''.

\textbf{NegaOctet.}
    The NegaOctet project~\cite{negaoctet} provides an illustration of traceability and maintenance issues in LCA data, used as background by many other studies. As a time‑limited research project, it did not guarantee long‑term data maintenance.
    According to \cite{negaoctet}, the database is no longer commercialized by the consortium. Only a fraction of it has been publicly disclosed and/or transferred to other databases (such as the French ADEME’s Base Empreinte). Based on the publicly available information to date, the current state of access to the data and knowledge contained in NegaOctet, as well as whether this data can be maintained over time, remains unclear. This situation raises concerns regarding downstream studies that reference this database, thus without possible access to the underlying data by the reader (e.g., a study by Capgemini relying on NegaOctet to estimate GPU impacts \cite{Desroches2025SustainableScalingAI}, as well as the recent A100 GPU study \cite{Falk2025MoreThanCarbon}, which specifies a database version of NegaOctet (2022) without indicating how the database can be accessed).\\
    \underline{Possible improvement}: \emph{(B3)} via \emph{(R1)} – ``Avoiding model washing'' (as for Ecologits),
    \emph{(B6)} via \emph{(R3)} – ``Connecting numbers back to models and data'', \emph{(B7)} via \emph{(R3)} – ``Supporting explanation, replication, and auditing'', \emph{(B8)} via \emph{(R4)} – ``Handling outdated components and evolution over time''.
%This situation illustrates how project‑based LCA data production can undermine traceability, reproducibility, and long‑term usability of environmental assessments.

These examples mirror potential model misuse: fragile, context‑ dependent models are reused as if they were robust, general laws, with limited visibility on their lineage, assumptions, and updates.

\subsection{Structural challenges in current LCA practice}
\label{sec_ex_LCA}

Beyond isolated anecdotes, several recurring structural issues in LCA practice make reduced rigor likely. \textcolor{red}{Some of them are direct instances of unmet requirements (e.g., \textit{R2} and the scope mismatches discussed below), which we illustrate below using concrete LCA references.}

\textbf{Sensitivity to database choice.}
    Impact results can substantially vary with the chosen background database, sometimes as much as with the impact assessment method itself.
    Recent work on GPUs \cite{Falk2025MoreThanCarbon} and comparative EPD studies \cite{Konradsen2024SameProductDifferentScore} show that changing only the background data can significantly alter conclusions.\\
    \underline{Possible improvement}: \emph{(B1)} via \emph{(R1)} - ``How much empirical support?''.

\textbf{Scope mismatches.}
    Secondary datasets often describe generic technologies, system boundaries, or use conditions that only partially match the studied product.
    Such mismatches are difficult to detect, yet they can affect the results when an “almost” relevant dataset is reused outside its original scope \cite{r740_lca, sanchez2022life, billaud2023ics, weppe2024streamlined}.
    \textcolor{red}{The paper \cite[Sec.~6]{Nordelof2014EnvironmentalImpactsElectrifiedVehicles} on the impacts of electric vehicles, carefully considering methodological flaws, also highlights the importance of time scope (related with the following ``changing technologies" item).} \\
    \underline{Possible improvement}: \emph{(B4)} via \emph{(R2, R4)} – ``Detecting incompatible assumptions'', \emph{(B5)} via \emph{(R2)} – ``Preventing out‑of‑domain reuse''.

\textbf{Non‑explainability of aggregated indicators.}
    LCA typically aggregates thousands of inventory flows into a small number of midpoint indicators.
    This many‑to‑few mapping is fundamentally non‑invertible: impact scores alone rarely allow experts to reconstruct underlying processes or assumptions.
    EPDs and background reports \cite{Springer_EPD_BackgroundReport_2025}, as well as recent AI‑related LCAs \cite{Elsworth2025GoogleServing,Schneider2025AIHardwareLCA}, exemplify how impact results are often published without the inventory or models needed for meaningful interpretation or reuse.\\
    \underline{Possible improvement}: \emph{(B6)} via \emph{(R3)} – ``Connecting numbers back to models and data'', \emph{(B7)} via \emph{(R3)} – ``Supporting explanation, replication, and auditing''.

\textbf{Integration barriers for research models.}
    Many academically proposed models are well documented but difficult to integrate into operational LCA workflows.
    They may be provided as stand‑alone scripts or spreadsheets \cite{Loubet2023ICT_HE,Zhang2022HV_AEC_LCA,Golard2024ParametricLCA,Golard2024PowerModel}, requiring substantial re‑implementation and interpretation before they can be composed with other models. This \emph{(i)} breaks the lineage or makes it very thin, \emph{(ii)} exposes to errors in the re-implementation and interpretation process \emph{(iii)} exposes one to "fantom obsolescence" if the original model reimplemented is found obsolete, among others.\\
    \underline{Possible improvement}: \emph{(B2)} via \emph{(R1)} – ``Where does uncertainty/ speculation enter?'', \emph{(B5)} via \emph{(R2)} – ``Preventing out‑of‑domain reuse''.

\textbf{Problematic proxies and changing technologies.}
    Simplifying proxies (e.g., scaling flash memory impacts by die area or capacity) may become invalid when technology changes, as with the transition from planar to 3D NAND \cite{WeppeNAND}. \textcolor{red}{See also \cite[Sec.~6]{Nordelof2014EnvironmentalImpactsElectrifiedVehicles} (time scope) for electrical vehicles.}
    Without explicit documentation of domains of validity, such proxies may continue to be applied long after their original assumptions are violated.\\
    \underline{Possible improvement}: \emph{(B2)} via \emph{(R1)} – ``Where does uncertainty/ speculation enter?'', \emph{(B4)} via \emph{(R2, R4)} – ``Detecting incompatible assumptions'', \emph{(B8)} via \emph{(R4)} – ``Handling outdated components and evolution over time''.

\textbf{Accounting conventions and truncation.}
    Multiple coexisting conventions for Scope~2 and Scope~3 emissions, or spend‑based versus activity‑based accounting, create additional layers of modeling choices.
    These conventions are often treated as interchangeable, even though they embed different system boundaries and truncation patterns. Convention choices can impact the whole chain: scope 3 emissions rely, almost by definition, on the models of the suppliers.\\
    \underline{Possible improvement}: \emph{(B4)} via \emph{(R2, R4)} – ``Detecting incompatible assumptions''.

\textbf{Correct interpretation of the impact results.}
\cite{Jacob2025InterpretingSustainabilityNumbers} highlights how sustainability indicators are often misinterpreted once detached from their methodological context, in particular through the frequent confusion between attributional and consequential (marginal) reasoning. Numbers derived from average, system‑level assessments are commonly used as if they represented the marginal impact of an additional use, leading to erroneous conclusions.
More generally, aggregated indicators are difficult for non‑experts to interpret, especially when they obscure whether impacts are localized or globally distributed.
In a similar vein, \cite{Finkbeiner2025MessageLCA} points an emerging gap between \emph{analysis‑LCA} and \emph{message‑LCA}, where communicated conclusions are ``partly in conflict with some key features and principles of LCA".
In the terms of our framework, such gaps amount to breaking the chain between models: LCA results are reused without preserving the lineage, scope, and uncertainty information encoded in the underlying models, so that the “message” may no longer reflects the conditions under which the analysis is valid. \\
\underline{Possible improvement}: \emph{(B5)} via \emph{(R2)} – ``Preventing out‑of‑domain reuse''.

\subsection{Why misuse is “natural”}
\label{sec_natural_model_misuse}

Model misuse rarely stems from bad intentions.
Early models are often developed under conditions of genuine ignorance: data are scarce, measurement is costly, and the main objective is to obtain an order‑of‑magnitude estimate that can guide first decisions. \textcolor{red}{This type of ignorance is commonly referred to as \textit{epistemic uncertainty}, which arises from incomplete knowledge and is, in principle, reducible, as opposed to \textit{aleatory uncertainty}, which reflects inherent variability.}
Engineers routinely navigate a precision–cost trade‑off: coarse models with few inputs are attractive when time or data are limited, whereas more detailed models demand extensive data collection and expertise.
Over time, however, the assumptions and scope of these initial models tend to be forgotten, especially when they are reused by people who were not involved in their design.

In LCA, this effect is amplified by the strong dependence on secondary data and by the diversity of goals and functional units.
The same physical system may be modeled differently for policy design, regulatory compliance, or product comparison, with different system boundaries and data requirements.
When results and models are transferred across these contexts without explicit scope and lineage information, methodological rigor degrades almost inevitably.

As highlighted by the benefits listed in Section~\ref{sec_req_model_lineage}, the proposed requirements and associated benefits can be read as a direct response to this natural drift and to the misuses illustrated above. The next section presents practical solutions to meet these requirements more consistently.

\section{Proposed framework}

The previous sections have identified four key requirements for more rigorous, explainable, and maintainable model‑based LCA: \emph{model lineage (R1)}, \emph{scope of models (R2)}, \emph{traceability (R3)}, and \emph{non‑obsolescence (R4)}.
We now outline a \textcolor{red}{computational-oriented} framework that operationalizes these requirements by \emph{(i)} structuring models as an explicit dependency graph \textit{(S1)}, \emph{(ii)} embedding them in an open, versioned repository with software‑like governance \textit{(S2)}, \emph{(iii)} automatic enforcement of integrity constraints \textit{(S3)}, and \emph{(iv)} establishing a well‑defined model taxonomy \textit{(S4)}.
\textcolor{red}{Figure~\ref{fig:lca-bipartite_solutions} illustrates the connections between the requirements and the proposed solutions.}

 \textcolor{red}{These principles are intended to serve as design guidelines of a concrete database implementation, such as ElecImpact \cite{ElecImpact_DEFIE}, a project of a new open and collaborative LCA database for electronics developed by the GDR DEFIE working group \cite{DEFIE_CNRS}.}

%Preventing model misusage: key elements for explainable and reproducible LCA \textcolor{red}{Or "Proposed solutions to rectify ICT LCA"}

%Recall of requirements:
%Model lineage  Section~\ref{sec_req_model_lineage} Traceability Section~\ref{sec_trac_scope} Scope of models Section~\ref{sec_scope_models} Non obsolescence Section~\ref{sec_model_maintenance}

\subsection{Model structure as a dependency graph}
\label{sec_model_structure}

A substantially more explicit model dependency graph is required than what is typically available today.
Such a graph should formally and explicitely represent models (and sub‑models) as nodes, and their dependencies as edges, making it possible to identify which results must be re‑evaluated when an upstream component is corrected, updated, or very importantly invalidated. Hence, 
systematic propagation of invalidation along this graph is essential for maintaining coherence across complex, compositional model ecosystems. This would mirror the logic of Common Vulnerabilities and Exposures (CVEs) used in software development. CVEs is a standardized system for identifying and cataloguing publicly known cybersecurity vulnerabilities \cite{CVEs}.

%POUR LE JOURNAL - voir si cet aspect de validation ne doit pas être déplacé, car il est un peu orthogonal à la model structure. Dans le sens que tous les modèles validés c'est pas suffisant si la structure de la composition n'est pas connue. Et à l'inverse, connaître parfaitement le graphe n'est pas suffisant si les modèles ne sont pas validés.
Ideally, a fully validated LCA model~$M$ would be confronted directly with measurements.
Since this is rarely feasible, a principled alternative is to require that the sub‑models used to construct~$M$ be individually validated within documented domains of validity and with quantified uncertainty.
The composite model~$M$ must then \emph{(i)} operate strictly within these domains, its \emph{context}, and \emph{(ii)} explicitly inherit and encode the assumptions and applicability criteria of its sub‑models.

This naturally leads to a hierarchy of models, where models can inherit from and be composed with one another.
More diffuse relationships, where a model is considered validated by analogy with another model, must also be made explicit.
Cycles in this dependency structure may reveal “incestuous” validation patterns with no solid empirical foundation; at minimum, such cycles should be identified and documented.

Concretely, each proposed model should document at least:
on which other models it depends, and whether these underlying models are validated and within which domains;
how its own domain of validity is defined (functional unit, system boundaries, technological scope, temporal and geographical coverage); how it can be validated or recalibrated if new data become available.
Proprietary or paid models that hide their internal structure and ancestry run counter to this vision, as they break dependency links and prevent tracing results back to empirical observations. 

\underline{This model structure supports requirements}: \emph{(R1)} model lineage, \emph{(R2)} scope of models, \emph{(R3)} traceability, and \emph{(R4)} non-obsolescence.

\subsection{Open and versioned repository}
\label{sec_open_and_versioned_repo}

Implementing such a dependency graph requires infrastructure similar to that used in modern software engineering.
To take another software development analogy: An LCA database can be compared to a compiled \emph{Docker image} or to a \emph{compiled binary}: a ready‑to‑use artifact encapsulating a complex model.
However, without the equivalent of a \emph{Dockerfile}, or of the \emph{source code}, that it, without an explicit recipe describing the model’s structure, assumptions, data sources, and parameterization, the model is not fully reproducible, inspectable, or explainable.

For instance, large software organizations increasingly rely on a \emph{monorepo} approach, where source code, dependencies, and build rules are maintained in a single, versioned repository \cite{monorepogoogle}.
Package managers (e.g. those used in Python, Linux distributions, or container ecosystems) then resolve dependencies, enforce compatibility constraints, and make it possible to reconstruct precise environments.

Another useful point of comparison can be found in open‑source software engineering practices, where the management of complex dependency structures has long been addressed through shared repositories, package managers, and strict versioning conventions. In the Python ecosystem, for instance, models and libraries are commonly developed in public version‑controlled repositories (e.g., GitHub) and distributed through centralized registries such as the Python Package Index (PyPI), with tools like pip enforcing explicit dependency declarations and compatibility constraints. Similar infrastructures exist in other communities, such as Maven Central for Java, where artifact versioning and dependency resolution are treated as first‑class concerns. Beyond their technical role, these ecosystems embody a culture of openness, reproducibility, and compatibility management, in which models are expected to declare their interfaces, assumptions, and supported versions. And as mentioned above, these ecosystems have in the last ~10 years integrated mechanism to track the CVEs across the web of dependencies.   

%We argue that LCA modeling would benefit from adopting a comparable culture, in which models are treated as reusable, versioned artifacts whose dependencies and domains of validity are explicitly managed, thereby supporting both long‑term maintainability and scientific reproducibility in open science.

LCA modeling should move toward a similar paradigm: a model repository in which every model is versioned, its ancestry is recorded, and its build instructions are part of the public record.
Such a repository, equipped with explicit versioning and dependency tracking, is the natural place to enforce the non‑obsolescence requirement \emph{(R4)} via efficient model maintenance: deprecated models remain archived and citable, but new versions can be linked, compared, and substituted in a controlled way.

%In addition to explicit dependency management and versioning, effective traceability also relies on the openness of models themselves. In this respect, traceability would be greatly facilitated by an open‑source culture in LCA model development.

\underline{This repository approach supports requirements}: \emph{(R1)} model lineage, \emph{(R3)} traceability, and \emph{(R4)} non-obsolescence.

\vspace{-1mm}
\subsection{Automatic enforcement of integrity\\ constraints}
\label{auto_enforc_inte_constraints}
\vspace{-1mm}

Integrity constraints, both across models and for compliance with mandatory methodology, should be enforced \emph{automatically}. %These constrained can be for instance mendatory product fields or admissible parent--child relations.
It mirrors integrity constraints in entity--relationship database schemas, and the way modern dependency managers verify and reconcile constraints to resolve dependency conflicts before producing a consistent build.

Typical checks could include the presence of mandatory \emph{product} and \emph{process} model parameters (see next subsection), adherence to allowed parent--child structures, conformance with relevant PCR, and schema-level assertions such as unit consistency, valid ranges, and type safety.

A pragmatic two-pass pipeline \textcolor{red}{may be} effective in practice. \emph{Pass~1} loads sources, instantiates models and data classes, populates a local database. \emph{Pass~2} aggregates and propagates constraints across the model graph, checks integrity constraints, and, on success, records standardized \emph{model validation}.

\underline{This automatic verification supports requirements}: \emph{(R3)} traceability and \emph{(R4)} non-obsolescence.

\vspace{-1mm}
\subsection{Well-defined model taxonomy}
\label{sec_model_cate}

In model‑based LCA, different types of models play distinct roles and may be categorized accordingly.
At a minimum, we think that the following categories are useful.

\textbf{Product and process models}, which specify typical technical parameters for a product or process category (e.g. a server product, PCB product, chip product, or assembly process, use phase process). \textcolor{red}{Some \textit{instances} of \textit{typed} nodes may claim average or generic configurations, while others may describe more specific cases. 
Similarly, defining subtypes of nodes enables finer granularity in the representation of products and processes while establishing a clear hierarchy.
As illustrated in Figure~\ref{fig:process}, products and processes can be represented as connected nodes, with the LCI forming a graph. Some nodes may be reused across multiple trees (i.e., across multiple LCIs).}

\textcolor{red}{\textbf{Impact models}, which compute environmental impacts for a product or process directly from product or process parameters, which may themselves be obtained through \emph{parameter conversion models}. As a result, these models may bypass the intermediate step of explicitly enumerating biosphere flows for the product under consideration. We distinguish three sub-classes, depending on whether parameters are mapped directly to environmental impact indicators (e.g., midpoint indicators), to biosphere-flow quantities, or to operational performance--resource proxies:}
\begin{itemize}
    \item \textcolor{red}{\textbf{Midpoint impact models}, which map product or process parameters directly to midpoint impact indicators, thereby bypassing explicit biosphere-flow modeling.
    \item \textcolor{red}{\textbf{Product-flow models}, which infer reference-flow quantities (and, more generally, biosphere-flow quantities) directly from technical parameters (e.g., mass from dimensions, number of wafers from die count).}}
    \item \textcolor{red}{\textbf{Computational handprint--footprint models}, which rely on operational proxy metrics, for instance, computational performance or complexity measures (e.g., GFLOPs) and energy-use indicators (e.g., Joules), to approximate, respectively, the service provided and its associated resource use.}
\end{itemize}

\textbf{Parameter conversion models}, which map technical characteristics of a product to other characteristics. These mappings may describe relationships between parent and child products, infer intra‑product characteristics (i.e., relationships between different parameters of the same product), or operate more generally between products and processes.

\textbf{Allocation models}, which distribute impacts or flows among co‑products, recycled materials, or services according to specified rules (e.g., module D‑type allocation, cut‑off or avoided burden rules, market‑based allocations for Scope~3).

\textcolor{red}{\textbf{Uncertainty models}, which infer probability distributions from input parameters to support stochastic analyses (e.g., Monte Carlo simulation). For example, the PEDIGREE method is an uncertainty model (see Section~\ref{sec_uncertainty_practice}).}

Making these categories explicit helps practitioners understand what each model does, how models can be composed, and where modeling choices enter the calculation.

For instance, a given product node in a tree graph can often be modeled in two complementary ways.
On one hand, it may be linked to an impact model that uses a small set of high‑level product parameters; on the other hand, it can be decomposed into a more detailed sub‑tree of sub-products and processes, each with their own parameters and models.
Within the same dependency graph, an analyst can either use the impact model or traverse the detailed sub‑tree to recompute impacts bottom‑up.
By making these alternative paths explicit, the framework exposes and documents the different approximations and assumptions.

\underline{This taxonomy supports requirements}: \emph{(R1)} model lineage and \emph{(R2)} Scope of models.

\vspace{-2mm}
\section{Conclusion}

A direct critique of current ICT service LCAs \cite{Coroama2020ICTServices} highlights the prevalence of simplistic assumptions, inadequate methodologies, and unsupported extrapolations. 
In this paper, we have complemented this perspective by arguing that LCA for ICT systems has effectively become a modeling exercise, in which complex systems of models replace direct observation of biosphere flows.
Because these models are composite, hierarchical, and only partially validated, ICT LCA requires an “extra‑high” level of carefulness in the way models are constructed, calibrated, integrated, and interpreted.
Our review of current practices suggests that this level of rigor is not always attained: assumptions and scopes may be under‑documented, dependencies may be opaque, and updates or corrections difficult to trace, so that apparently precise impact figures may rest on fragile foundations.

From emblematic examples and structural challenges, we have distilled four key requirements for credible model‑based LCA in ICT: explicit model lineage, clearly defined model scope, end‑to‑end traceability, and managed non‑obsolescence.
We then outlined a framework that operationalizes these requirements by representing models as explicit dependency graphs, embedding them in an open, versioned LCA model repository, automatically enforcing integrity constraints, and establishing a well-defined model taxonomy.

\textcolor{red}{More broadly, verification tools are also being developed in other scientific fields, for instance, to detect problematic publications and mitigate their downstream effects (e.g., biases in meta-analyses) through \textit{living systematic reviews} \cite{Cabanac2022ProblematicPaperScreener,GranaPossamai2025RetractedMetaAnalyses,COVIDNMA}, as well as to enable computer-assisted formal verification via proof assistants such as Lean or Rocq, which have already revealed flaws in widely cited results \cite{ToobySmith2026Formalizing2HDM,Sparkes2026ComputerFlawPhysics}.
}

In this respect, adopting a skeptical attitude of the kind advocated by Richard Feynman \cite{Feynman1985Surely}, critical questioning, transparency, and a constant effort to distinguish robust results from convenient but fragile claims, %together with sharing models, documenting their lineage and scope, and treating databases as evolving, auditable software artefacts, 
are concrete steps toward reliable ICT LCAs.

\bibliographystyle{ACM-Reference-Format}
\bibliography{references}

\end{document}